# Efficient target control of complex networks based on preferential matching

Xizhe Zhang [1†], Huaizhen Wang[1], Tianyang Lv [2,3]

[1] （School of Computer Science and Engineering, Northeastern University, Shenyang110819, China）

[2] （College of Computer Science and Technology, Harbin Engineering University, Harbin 150001, China)

[3] （IT Center, National Audit Office, Beijing 100830, China)

Email: zhangxizhe@mail.neu.edu.cn

**Abstract:** Controlling a complex network towards a desire state is of great importance in many applications. Existing works present an approximate algorithm to find the driver nodes used to control partial nodes of the network. However, the driver nodes obtained by this algorithm depend on the matching order of nodes and cannot get the optimum results. Here we present a novel algorithm to find the driver nodes for target control based on preferential matching. The algorithm elaborately arrange the matching order of nodes in order to minimize the size of the driver nodes set. The results on both synthetic and real networks indicate that the performance of proposed algorithm are better than the previous one. The algorithm may have various application in controlling complex networks.

**Introduction**

Controlling a complex networked system play an important role in many nature and technology regions. According to the control theory [1-3], a system is controllable if it can be driven from any initial state to any desired state in the finite time. The nodes used to input the external signals are called driver nodes [4].

The driver nodes used to fully control the network can be obtained by maximum matching of the network [4], the unmatched nodes are minimum set of driver nodes (in short MDS). Based on this framework, lots of works analyzed the structural properties of MDS [5-8], roles of nodes in control [9], and robustness of controllability [10]. However, in many real control scenario, only a small fraction of nodes need to be controlled, which are called target control [11]. In order to find the driver nodes to control specific target nodes, recent works [11] presented an analysis framework to investigate the target control of complex network. They proposed an approximate greedy algorithm (*GA*) based on multiple maximum matchings to obtain the driver nodes used to control the target nodes.

However, the *GA* algorithm can only find the approximate minimum driver nodes set. If there exist more than one maximum matchings in the networks, the results of *GA* algorithm strongly depend on what maximum matchings are selected. Therefore, the number of driver nodes may varies a large range [11] (Figure.1). How to find the minimum number of driver node for target control is still an unsolved problem.

Here we presented a novel algorithm for finding driver nodes to control target nodes of network. In contrast to previous approach, we elaborately arranged the matching order of nodes and minimized the total number of result driver nodes. The results on both synthetic and real networks show that the driver nodes we obtained are less than *GA* algorithm.

## Method

For a network $G(V, E)$ and a target nodes set $T \in V$, we say the system is target controllable if the state of the target nodes can be driven from any initial state to desire final state. Previous work [11] proposed a $k$-walk theory and proved that in a tree-like network (in which there are no loop), if a node has paths with different distance to each target node, then the node can control these target nodes. However, single node cannot control all target nodes in many network. Therefore, previous network [11] proposed an approximate algorithm based on multiply maximum matchings to obtain the driver nodes. The algorithm first construct a series of bipartite graphs, and then find the maximum matchings of each bipartite graph. The union of unmatched nodes of all bipartite graph are the driver nodes used to control target nodes. (Examples are shown in Figure.1).

The key idea of above algorithm is to find maximum matching for each sub-bipartite graph. However, the maximum matchings of a network are not unique in most network. Therefore, even for a simple network, the algorithm would produce different results with different maximum matchings. For example, for network shown in Figure.1A, the algorithm will obtain three different driver nodes sets, which are $D_1 = \{1, 2, 5\}$, $D_2 = \{1, 5\}$ and $D_3 = \{1\}$. The reason for multi-results are the maximum matchings used in the algorithm are different. For example, If we match edge $e(1\rightarrow 4)$ rather than $e(2\rightarrow 4)$ in sub-bipartite graph 3, the node 2 will not act as driver nodes, and the driver nodes set will be $D_2 = \{1, 5\}$. If we match edge $e(6\rightarrow 7)$ rather than $e(5\rightarrow 7)$ in sub-bipartite graph 2, we will have only one driver node $D_1=\{1\}$ to control entire target nodes set.

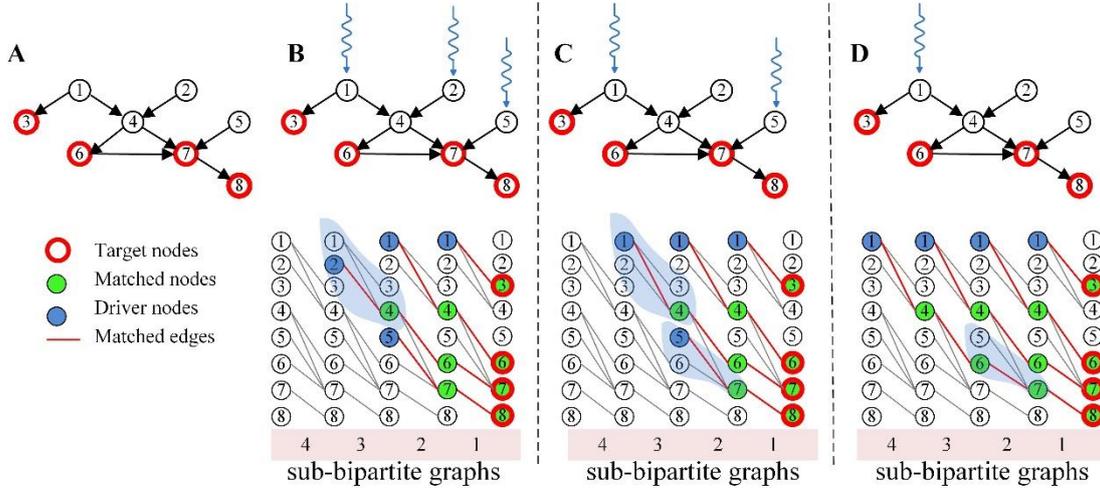

**Figure.1** | Illustration of random matching of GA algorithm. (A) A sample network $G$ with target nodes $\{3, 6, 7, 8\}$; (B-D). Three DSs obtained by GA algorithm and their matching process, in which $D_1 = \{1, 2, 5\}$, $D_2 = \{1, 5\}$, $D_3 = \{1\}$. The driver nodes are obtained by the following process: 1. Construct a bipartite graph $B$ (sub-bipartite graph 1) which the right side contain all target nodes and the left side contain the nodes pointing to the target nodes; 2. Find a maximum matching of $B$ and let the matched nodes be $M$; 3. Let the $M$ be the new target nodes and repeat the step 1 and 2 until no new matched nodes founded. The difference of three matching processes are highlight by blue shadow.

Therefore, to minimize the number of total driver nodes, we need to select the appropriate maximum matching for each sub-bipartite graph. However, the number of unmatched nodes of each sub-bipartite graph are fixed because the maximum matchings of each bipartite graph have same size. The only way to decrease the number of driver nodes is to make the driver nodes of different sub-bipartite graphs be overlapped with each other, and the total number of driver nodes will be decreased. For example, in Figure.1D, the unmatched node of all four sub-bipartite graphs is node 1, which make the total number of driver nodes decrease from three to one.

In order to assure more driver nodes of current sub-bipartite graph overlapped with that of previous sub-bipartite graphs, we use the preferential matching [12] to obtain the expected driver nodes. This method match the nodes according pre-defined order, and guarantee the nodes in the rear of the queue to be driver nodes. Therefore, the problems are how to choose the appropriate driver nodes of each sub-bipartite graphs in order to minimize the total number of driver nodes. Here we presented the following strategies:

1. The driver nodes of current sub-bipartite graph should be overlapped with driver nodes of the previous sub-bipartite graph, that will decrease the total number of driver nodes;
2. The nodes which frequently appear in the matching graph (all sub-bipartite graphs) should be driver nodes with high priority, which will give the nodes in the following sub-bipartite graphs with high probability to overlap with existed driver nodes.

Figure.2 shows an example network about above strategies. For network shown in Figure.2A, we construct a matching graph (*MG*) which starts from the target nodes, and iteratively add the parent nodes of current nodes to the graph, until no nodes are added. We count the frequency of each node appeared in the matching graph and arrange the nodes ascendingly by the frequency. For example, Figure.2B shown the matching graph of Figure.2A, and the count of nodes 1-8 are {4, 3, 1, 3, 2, 3, 2, 1}. Therefore, the matching sequence should be nodes {8, 3, 7, 5, 4, 6, 2, 1}. For each sub-bipartite graphs of *MG*, we used the above matching sequence to find driver nodes.

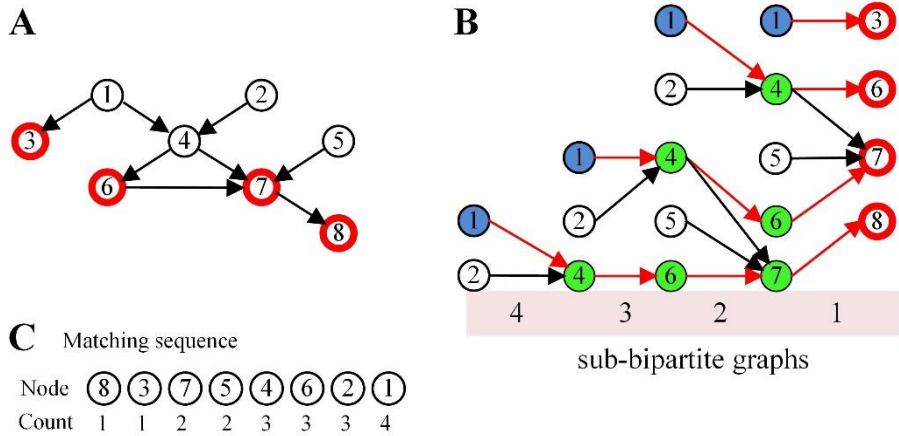

**Figure.2** | Illustration of preferential matching for target control. (A). A sample network *G* with target nodes {3, 6, 7, 8}. (B). Matching graph for target nodes {3, 6, 7, 8}. (C). Matching sequence of nodes based on their counts in the matching graph. The counts for nodes sequence {1,2,6,4,5,7,3,8} are {4,3,3,3,2,2,1,1}.

Overall, for a network *G* and target nodes set *T*, the algorithm based on preferential matching (*PM*) for finding driver nodes is listed as follows:

1. For target nodes set *T*, construct bipartite graph $B_1(F, T)$, where *F* are the nodes set pointing to target nodes set *T*;
2. Let *F* be the new target nodes set, repeat step 1 to construct bipartite graph $B_i(F, T)$ until no nodes founded. Let the matching graph $M(T)=\{B_1, B_2, …, B_i\}$;
3. For each nodes in *M*, compute their counts *f(n)*, arrange the nodes by ascending order of *f(n)* and let the matching sequence be *Q*;

4. For sub-bipartite graph of *M*, find the maximum matching based on preferential matching by using node sequence *Q*. Let the driver nodes be $D=\{d_1,d_2,…,d_i\}$; rearrange the node sequence by putting the nodes of *D* in the rear of *Q*;
5. Repeat step 4, find driver nodes $D_i$ of sub-bipartite graph $B_i$; the final driver nodes to control target nodes are $D=\bigcup D_i$;

**Result**

To quantify the efficiency of the algorithm, we evaluated the fraction of driver nodes $n_D=N_D/N$ based on preferential matching (*PM*) algorithm and greedy algorithm (*GA*) algorithm [10]. We used the following two different schemes for target nodes selection:
1. Random selection scheme: Random select nodes uniformly from the network until it reach expected target fraction *f*;
2. Local selection scheme: Random select a seed node, expand it based on *breadth-first search* (BFS) tree until it reach expected target fraction *f*.

Figure 3 show the results of two scale-free networks [13] with $N=10^4$. For different target node fraction $f \in [0,1]$, the *PM* algorithm always has better performance than *GA* algorithm in both two target nodes selection schemes. Furthermore, the difference of $n_D$ obtained by *PM* and *GA* algorithm $|\Delta n_D|=|n_{D\text{-}GA}- n_{D\text{-}PM}|$ are increasing with the fraction of target nodes *f*, suggesting the *PM* algorithm is more efficient in controlling large fraction of target nodes.

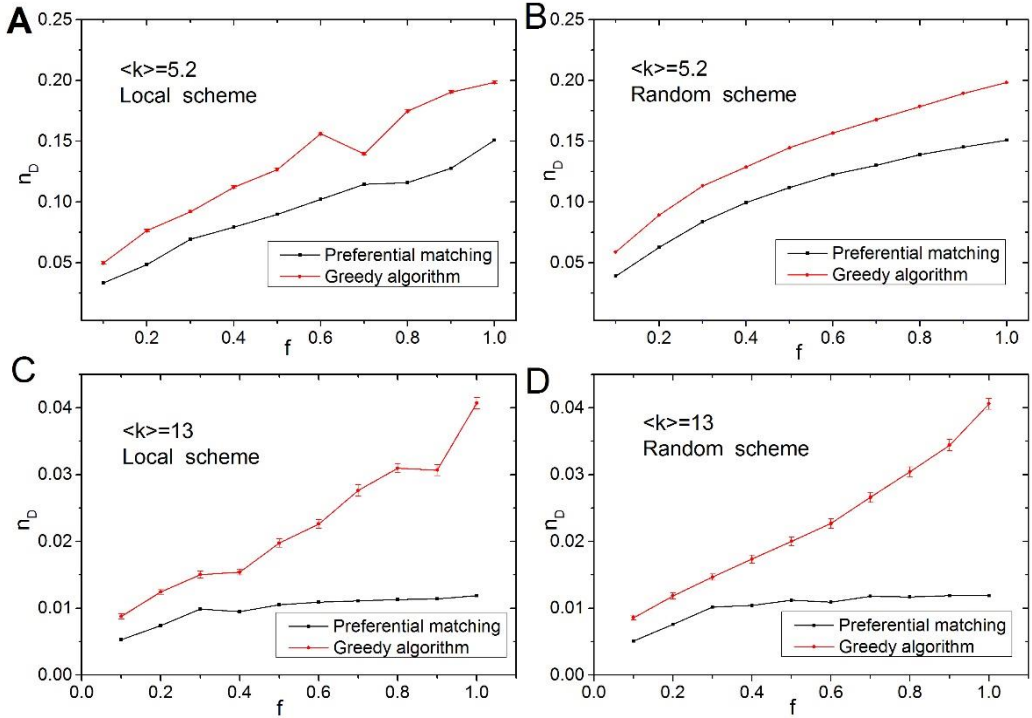

**Figure.3** | Efficiency analysis of target control algorithm for two synthetic networks (A-B). For scale-free network with $N=10^4$ and $<k>=5.2$, we show the density of driver nodes as the function of fraction of target nodes. (A) Result of local selection scheme; (B) Results of random selection scheme; (C-D). For scale-free network with $N=10^4$ and $<k>=13$, we show density of driver nodes as the function of fraction of target nodes. (C) Result of local selection scheme; (D) Results of random selection scheme. For each network, we compute the fraction of driver nodes $n_D$ based on

preferential matching and greedy algorithm. For greedy algorithm, the $n_D$ are computed based on the results of 100 random experiments.

Next we analyzed the $n_D$ with different average degree $<k>$. Figure.4 show the results for both Scale-free networks and *ER* random network based on local and random target selection schemes. The *PM* algorithm have lower $n_D$ in all networks than *GA* algorithm. Note that the variation of $n_D$ of local selection scheme of target nodes are much larger than random selection scheme, suggest that there are many driver nodes set to control target nodes which local connected.

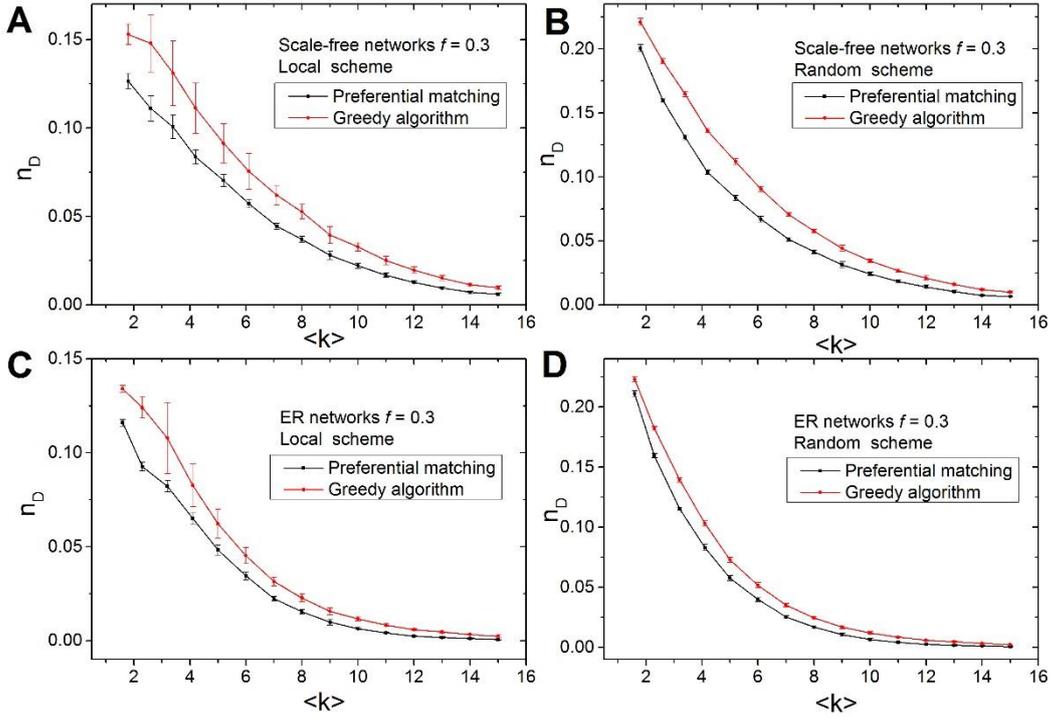

**Figure.4** | Efficiency of the algorithm for different average degree $<k>$. (A-B). For scale-free network with $N=10^4$ and target nodes fraction $f=0.3$, we show (A) density of driver nodes versus $<k>$ based on local selection scheme, and (B) density of driver nodes versus $<k>$ based on random selection scheme; (C-D). For ER random network with $N=10^4$ and target nodes fraction $f=0.3$, we show (C) density of driver nodes versus $<k>$ based on local selection scheme, and (D) density of driver nodes versus $<k>$ based on random selection scheme.

We also evaluated the performance of the *PM* algorithm in real networks. The networks are selected based on their diversity of topological structure, including the food web networks, Transcription networks, citation network, Internet and et.al. Table.1 and Figure.5 shows the results. For all networks and the fraction of target nodes in both random and local schemes, the *PM* algorithm always has better performance than *GA* algorithm.

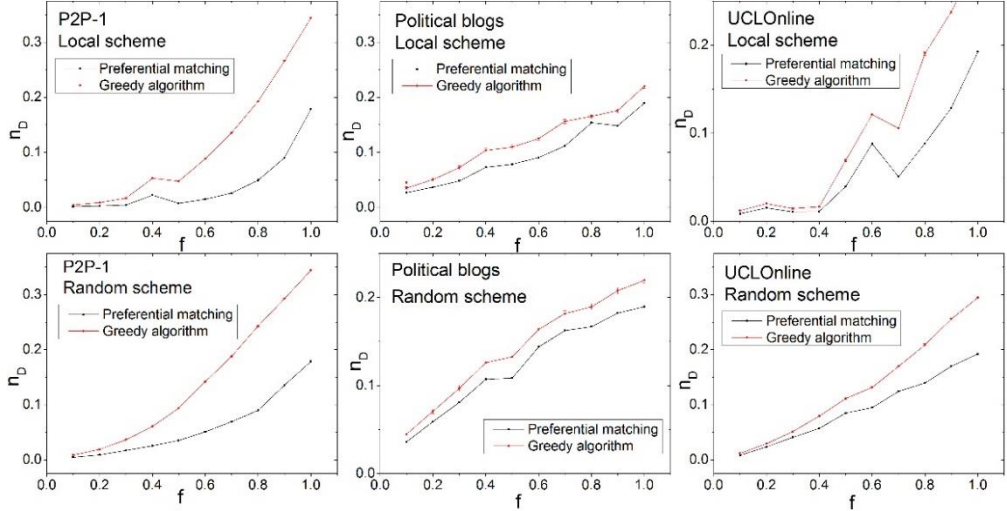

**Figure.5** | Result of real networks. We show the fraction of driver nodes and the fraction of target nodes. The PM method always have better performance in both local and random target nodes selection scheme.

**Table1** | Results of the real networks analyzed in the paper. For each network, we show its type, name, number of nodes (*N*) and edges (*L*), average degree <*k*>, density of driver nodes ($n_{pd}$) based on preferential matching, density of driver nodes $n_{rd}$ based on random matching. The fraction of target nodes *f*=0.5.

| Type | Name | *N* | *L* | <*k*> | Random selection | | Local selection | |
|---|---|---|---|---|---|---|---|---|
| | | | | | $n_{pd}$ | $n_{rd}$ | $n_{pd}$ | $n_{rd}$ |
| Food Web | Mangrove[14] | 97 | 1492 | 30.76 | 3.09% | 5.05% | 4.12% | 6.27% |
| | Silwood[15] | 154 | 370 | 4.81 | 30.52% | 31.27% | 24.03% | 25.37% |
| Neuronal | C. elegans[16] | 306 | 2345 | 15.33 | 7.07% | 7.97% | 4.38% | 5.55% |
| Transcription | E.Coli[17] | 423 | 578 | 2.73 | 39.95% | 43.04% | 32.39% | 38.02 |
| | TRN-Yeast-1[18] | 4441 | 12873 | 5.80 | 45.49% | 45.68% | 43.75% | 44.13% |
| | TRN-Yeast-2[19] | 688 | 1079 | 3.14 | 39.68% | 40.67% | 31.40% | 33.33% |
| Trust | Prison inmate[20, 21] | 67 | 182 | 5.43 | 13.43% | 17.51% | 11.94% | 13.84% |
| | WikiVote[22] | 7115 | 103689 | 29.15 | 39.68% | 40.33% | 37.26% | 38.37% |
| Electronic circuits | s208a[23] | 122 | 189 | 3.10 | 9.02% | 14.75% | 6.56% | 10.03% |
| | s420a[23] | 252 | 399 | 3.17 | 8.33% | 14.43% | 3.57% | 7.33% |
| | s838a[23] | 512 | 819 | 3.20 | 6.64% | 12.10% | 1.95% | 5.77% |
| Citation | ArXiv-HepTh[24] | 27770 | 352807 | 25.41 | 13.41% | 15.10% | 7.62% | 10.03% |
| | Kohonen[25] | 4470 | 12731 | 5.70 | 17.90% | 21.11% | 9.49% | 13.26% |
| WWW | Political blogs[26] | 1224 | 16718 | 27.32 | 10.87% | 13.26% | 7.84% | 10.99% |
| Internet | p2p-1[27] | 10876 | 39994 | 7.35 | 3.51% | 9.41% | 0.77% | 4.80% |
| | p2p-2[27] | 8846 | 31839 | 7.20 | 4.70% | 10.44% | 2.32% | 6.11% |
| | p2p-3[27] | 8717 | 31525 | 7.23 | 4.73% | 10.52% | 2.44% | 7.48% |
| Social network | UClonline[28] | 1899 | 20296 | 21.38 | 8.48% | 11.14% | 3.95% | 6.91% |
| | Facebook_0[29] | 347 | 5038 | 29.04 | 1.80% | 2.69% | 0.30% | 0.60% |
| | Facebook_107[29] | 1912 | 53498 | 55.96 | 0.10% | 0.19% | 0.10% | 0.10% |
| | Facebook_348[29] | 572 | 6384 | 22.32 | 0.89% | 1.79% | 0.45% | 0.45% |

## Discussion

The controllability of complex network is of great importance in many applications. How to control a small fraction of target nodes is a common task in many real control scenarios. Here we proposed a novel algorithm based on preferential matching to minimize the number of driver nodes. The main processes of our algorithm are the same as the previous one [11] based on multi maximum matching of the induced bipartite graphs. However, we elaborately arranged the matching order of the nodes, which can significantly reduce the number of the result driver nodes.

However, our algorithm still cannot guarantee the result are the optimum one. The future works should focus on finding an efficient and precise method to minimize the driver nodes.

## Reference

1. Kalman, R.E., *Mathematical Description of Linear Dynamical Systems.* Journal of the Society for Industrial & Applied Mathematics, 1963. **1**(2): p. 152-192.
2. Luenberger, D.G., *Introduction to Dynamic Systems: Theory, Models, & Applications.* Proceedings of the IEEE, 1979. **69**(9): p. 1173.
3. Slotine, J.J.E. and W. Li, *Applied nonlinear control*2004: China Machine Press.
4. Liu, Y.Y., J.J. Slotine, and A.L. Barabasi, *Controllability of complex networks.* Nature, 2011. **473**(7346): p. 167-73.
5. Jia, T. and A.L. Barabasi, *Control capacity and a random sampling method in exploring controllability of complex networks.* Sci Rep, 2013. **3**: p. 2354.
6. Jia, T. and M. Posfai, *Connecting core percolation and controllability of complex networks.* Sci Rep, 2014. **4**: p. 5379.
7. Posfai, M., et al., *Effect of correlations on network controllability.* Sci. Rep., 2013. **3**.
8. Piao, X., et al., *Strategy for community control of complex networks.* Physica A Statistical Mechanics & Its Applications, 2015. **421**: p. 98-108.
9. Ruths, J. and D. Ruths, *Control profiles of complex networks.* Science, 2014. **343**(6177): p. 1373-6.
10. Pu, C.L., W.J. Pei, and A. Michaelson, *Robustness analysis of network controllability.* Physica A Statistical Mechanics & Its Applications, 2012. **391**(18): p. 4420-4425.
11. Gao, J., et al., *Target control of complex networks.* Nat Commun, 2014. **5**: p. 5415.
12. Zhang, X., et al., *Structural controllability of complex networks based on preferential matching.* PLoS One, 2014. **9**(11): p. e112039.
13. Barabasi, A.L. and R. Albert, *Emergence of scaling in random networks.* Science, 1999. **286**(5439): p. 509-12.
14. Ulanowicz, R.E. and D.L. DeAngelis, *Network analysis of trophic dynamics in south florida ecosystems.* US Geological Survey Program on the South Florida Ecosystem, 2005. **114**.
15. Albert, R. and A. Barabási, *Statistical mechanics of complex networks.* Lecture Notes in Physics, 2001. **74**(1): p. xii.
16. Watts, D.J. and S.H. Strogatz, *Collective dynamics of 'small-world' networks.* nature, 1998. **393**(6684): p. 440-442.
17. Shen-Orr, S.S., et al., *Network motifs in the transcriptional regulation network of Escherichia coli.* Nature Genetics, 2002. **31**(1): p. 64-8.
18. Bu, D., et al., *Topological structure analysis of the protein–protein interaction network in budding yeast.* Nucleic acids research, 2003. **31**(9): p. 2443-2450.

19. Milo, R., et al., *Network motifs: Simple building blocks of complex networks.* Science, 2002. **42**(6821): p. 285-298.
20. Milo, R., et al., *Superfamilies of evolved and designed networks.* Science, 2004. **303**(5663): p. 1538-1542.
21. Van Duijn, M.A., et al., *Evolution of sociology freshmen into a friendship network.* Journal of Mathematical Sociology, 2003. **27**(2-3): p. 153-191.
22. Leskovec, J., et al., *Community structure in large networks: Natural cluster sizes and the absence of large well-defined clusters.* Internet Mathematics, 2009. **6**(1): p. 29-123.
23. Milo, R., et al., *Network motifs: simple building blocks of complex networks.* Science, 2002. **298**(5594): p. 824-827.
24. Leskovec, J., J. Kleinberg, and C. Faloutsos. *Graphs over time: densification laws, shrinking diameters and possible explanations*. in *Proceedings of the eleventh ACM SIGKDD international conference on Knowledge discovery in data mining*. 2005. ACM.
25. M. S. Handcock, D.H., C. T. Butts, S. M. Goodreau, M. Morris, . *Statnet: An R package for the Statistical Modeling of Social Networks*. 2003. Available from: http://www.csde.washington.edu/statnet.
26. Adamic, L.A. and N. Glance. *The political blogosphere and the 2004 US election: divided they blog*. in *Proceedings of the 3rd international workshop on Link discovery*. 2005. ACM.
27. Leskovec, J., J. Kleinberg, and C. Faloutsos, *Graph evolution: Densification and shrinking diameters.* ACM Transactions on Knowledge Discovery from Data (TKDD), 2007. **1**(1): p. 2.
28. Opsahl, T. and P. Panzarasa, *Clustering in weighted networks.* Social Networks, 2009. **31**(2): p. 155-163.
29. Mcauley, J.J. and J. Leskovec, *Learning to discover social circles in ego networks.* Advances in Neural Information Processing Systems, 2012: p. 539-547.